\documentclass[a4paper, 11pt]{article}
\usepackage[T1]{fontenc}
\usepackage{jheppub,graphicx,epsf,amssymb,amsbsy,amsfonts,amssymb,amsmath, empheq}
\def\be{\begin{equation}}
\def\ee{\end{equation}}
\def\bea{\begin{eqnarray}}
\def\eea{\end{eqnarray}}

\def\bg{\bar{g}}
\def\beq{\begin{eqnarray}}\def\eeq{\end{eqnarray}}
\def\ba#1\ea{\begin{align}#1\end{align}}
\def\bg#1\eg{\begin{gather}#1\end{gather}}
\def\bm#1\em{\begin{multline}#1\end{multline}}
\def\bmd#1\emd{\begin{multlined}#1\end{multlined}}

\def\D{\Delta}

\def\({\left(}
\def\){\right)}
\def\[{\left[}
\def\]{\right]}

\def\D{\Delta}

\title{Subsubleading soft graviton symmetry and MHV graviton scattering amplitudes}

\author[1,2]{Shamik Banerjee}
\author[3]{Sudip Ghosh}
\author[4,5]{Sai Satyam Samal}
\affiliation[1]{Institute of Physics, Sachivalaya Marg, Bhubaneshwar, India-751005}
\affiliation[2,5]{Homi Bhabha National Institute, Anushakti Nagar, Mumbai, India-400085}
\affiliation[3]{Okinawa Institute of Science and Technology,1919-1 Tancha, Onna-son, Okinawa 904-0495,Japan}
\affiliation[4]{National Institute of Science Education and Research, Bhubaneswar 752050, Odisha, India}
\emailAdd{banerjeeshamik.phy@gmail.com, sudip112phys@gmail.com, saisatyam.phy@gmail.com}

\abstract{In arXiv:2008.04330 it was shown that supertranslation and $\overline{SL(2,\mathbb C)}$ current algebra symmetries, corresponding to leading and subleading soft graviton theorems, are enough to determine the tree level MHV graviton scattering amplitudes. In this note we clarify the role of subsubleading soft graviton theorem in this context.}

\begin{document}
\maketitle
\flushbottom

\section{Introduction}
Celestial holography is now an active area of research. In short celestial holography \cite{Pasterski:2016qvg,Arkani-Hamed:2020gyp} is the statement that the holographic dual of quantum theory of gravity in asymptotically flat four dimensional space time is a CFT \cite{Kapec:2014opa} living on the two dimensional celestial sphere. We do not know of any Lagrangian description of the dual CFT but owing to the infinite dimensional symmetries \cite{Sachs:1962zza,Strominger:2013lka,Strominger:2013jfa,Barnich:2009se,Barnich:2011ct,Banerjee:2020zlg,Banerjee:2020vnt,Donnay:2020guq,Guevara:2021abz} we can solve the CFT in certain special cases. Explicit examples of this appeared in \cite{Banerjee:2020zlg,Banerjee:2020vnt} where the authors could determine the tree level MHV graviton and gluon scattering amplitudes in Einstein gravity and pure Yang-MIlls theory respectively using the infinite dimensional asymptotic symmetries. In \cite{Banerjee:2020zlg} it was shown that the subleading soft graviton theorem \cite{Cachazo:2014fwa} is the same as the Ward identity of the $\overline{SL(2,\mathbb C)}$ current algebra. It turns out \cite{Banerjee:2020zlg} that the $\overline{SL(2,\mathbb C)}$ current algebra symmetry together with the supertranslations \cite{He} are enough to fix the MHV graviton scattering amplitudes. 

The question that remains open is what happens to the subsubleading soft graviton symmetry. For example, global subsubleading soft graviton symmetry has been used in \cite{Pate:2019lpp} to determine the leading term in the celestial OPE \cite{Pate:2019lpp, Banerjee:2020kaa} of two graviton primaries. In this paper we clarify the role of the subsubleading soft symmetry in the MHV sector. In particular, we are interested here in understanding to what extent the subsubleading soft symmetry is required to determine the full tree-level MHV scattering amplitude instead of just the leading OPE coefficient in the celestial OPE. 

In \cite{Banerjee:2020zlg}, null states involving $\overline{SL(2,\mathbb C)}$ current algebra and supertranslation descendants were found in the MHV sector. A brief review of the implications of these null states for MHV amplitudes is presented in Section \ref{review}. In this paper we show that new null states involving subsubleading current algebra descendants appear as a result of consistency between the OPE of two graviton primaries in the MHV sector and the subsubleading soft graviton theorem. These new null states can also be determined by applying the global subsubleading symmetry to the above mentioned supertranslation and $\overline{SL(2,\mathbb C)}$ current algebra null states \cite{Banerjee:2020zlg} which appear in the MHV sector.  So, the new null states involving subsubleading current algebra descendants do not carry any additional information. We also show that the OPE itself, including descendant contributions, is invariant under the subsubleading current algebra symmetry, as expected. 

Recently in \cite{Guevara:2021abz}, the algebra of an infinite tower of conformal soft currents has been derived. There it was noted that the additional tower of currents beyond the ones associated to the leading, subleading and subsubleading soft graviton theorems do not yield any further constraints on the S-matrix. We would like to emphasise however that in our case of interest, it is sufficient to consider the constraints due to the supertranslation and $\overline{SL(2,\mathbb C)}$ current algebra symmetries, which arise from the leading and subleading soft graviton theorems respectively, for determining tree-level MHV graviton amplitudes in Einstein gravity. 

This paper is structured as follows. In Section \ref{review} we review the main results of \cite{Banerjee:2020zlg} regarding how MHV graviton amplitudes are determined by supetranslations and $\overline{SL(2,\mathbb C)}$ current algebra symmetries associated to the leading and subleading positive helicity soft graviton theorems respectively. In Section \ref{subsubsoft}, the subsubleading soft graviton symmetry is discussed. In Section \ref{subsubrels1}, we derive linear relations among descendants created by action of subsubleading current algebra, supertranslations and $\overline{SL(2,\mathbb C)}$  current algebra generators by demanding consistency of the OPE in the MHV sector with the subsubleading conformal soft theorem. These relations are then re-derived in Section \ref{subsubrels2} from the null states of \cite{Banerjee:2020zlg} using the global subsubleading soft symmetry. In Section \ref{subsubope1}, the invariance of the first subleading term in the OPE between positive helicity gravitons under the global subsubleading symmetry is checked. Appendix  \ref{comms} collects various relevant commutation relations involving symmetry generators appearing in this paper. Appendix \ref{subsubrels3} contains the derivation of another null state involving a subsubleading current algebra descendant. Finally in Appendix \ref{opeinvcheck}, further subleading terms in the positive helicity graviton OPE are shown to be invariant under the global subsubleading symmetry.

\section{Review}
\label{review}

In this section we briefly review the main results of \cite{Banerjee:2020zlg} regarding the constraints on the MHV sector imposed by supertranslations and $\overline{SL(2,\mathbb C)}$ current algebra which are associated with the leading and subleading positive helicity soft graviton theorems respectively. 

\subsection{MHV sector of Celestial CFT}

In \cite{Banerjee:2020zlg}, the primary objects of study were tree-level MHV graviton scattering amplitudes in Einstein gravity. A crucial feature of such amplitudes is that they are closed under soft limits. This is because if we take the soft limit of a positive helicity graviton in a MHV amplitude, then soft theorems relate this to a lower point MHV amplitude. However, making a negative helicity graviton soft yields a lower point amplitude where only one graviton has negative helicity and the rest have positive helicity. Such amplitudes vanish at tree-level in Einstein gravity. This implies that negative helicity soft gravitons decouple from the MHV sector. The significance of this decoupling is that it allows one to restrict attention to the symmetries generated by positive helicity soft gravitons only. This sector is also closed under taking collinear limits. This implies closure of the OPEs of graviton primaries in the corresponding sector of the dual Celestial CFT (CCFT). 

The above properties of MHV amplitudes enable us to define an autonomous MHV sector of the CCFT which provides a holographic description of tree-level MHV graviton scattering amplitudes in Einstein gravity. 

\subsection{Symmetries of graviton scattering amplitudes : Supertranslations and $\overline{SL(2,\mathbb C)}$ current algebra}




In CCFT, the currents that generate symmetries of the S-matrix correspond to conformal primary operators whose scaling dimension $\Delta$ take particular integer values. Such operators are referred to as conformal soft operators.  The Ward identities for these currents are given by conformal soft theorems which are simply the usual energetic soft theorems expressed in the conformal basis \cite{Donnay:2018neh}. In the analysis of \cite{Banerjee:2020zlg}, positive helicity conformal soft graviton operators with $\Delta=1$ (leading conformal soft) and $\Delta=0$ (subleading conformal soft) played a central role. The conformal soft theorems associated to these operators were shown in \cite{Banerjee:2020zlg} to be equivalent to the Ward identities for supertranslations and a $\overline{SL(2,\mathbb{C})}$ current algebra respectively. Let us now review these symmetries.

\subsubsection{Leading soft graviton and supertranslations}

Let $G^{+}_{\Delta}(z,\bar{z})$ be a positive helicity graviton primary with scaling dimension $\Delta$ inserted on the $2$-$d$ celestial sphere at $(z,\bar{z})$. The leading conformal soft positive helicity graviton is then defined as
\begin{equation}
\label{Sp0}
\begin{split}
S_{0}^{+}(z,\bar{z}) = \lim_{\Delta \to 1} (\Delta -1) G^{+}_{\Delta}(z,\bar{z}) 
\end{split}
\end{equation}

The operator $S_{0}^{+}(z,\bar{z})$ can be expanded as \cite{Banerjee:2020zlg}
\begin{equation}
\label{P0Pm1}
\begin{split}
S_{0}^{+}(z,\bar{z}) =  P_{0}(z) + \bar{z} P_{-1}(z)
\end{split}
\end{equation}


where $P_{0}(z)$ and $P_{-1}(z)$ are currents whose Ward identities are determined by the leading conformal soft graviton theorem. Let us define the modes of these currents on the celestial sphere as $P_{n,0}$ and $P_{n,-1}$ where $n \in \mathbb{Z}$. Their algebra is given by
\begin{equation}
\label{PPcom}
\begin{split}
[P_{n,m}, P_{n',m'}] = 0
\end{split}
\end{equation}

where  $m,m'= -1,0$. Acting on a conformal primary $\mathcal{O}_{h,\bar{h}}(z,\bar{z})$ these modes generate supertranslations 
\begin{equation}
\label{suptrans}
\begin{split}
[P_{n,m}, \mathcal{O}_{h,\bar{h}}(z,\bar{z})] = \epsilon z^{n+1} \bar{z}^{m+1} \mathcal{O}_{h+1/2,\bar{h}+1/2}(z,\bar{z})
\end{split}
\end{equation}

where $\epsilon =\pm 1$ if $\mathcal{O}_{h,\bar{h}}$ denotes an outgoing (incoming) state in the S-matrix. Note that $4$-$d$ global spacetime translations are generated by the set $\{ P_{-1,0}, P_{0,0}, P_{0,-1}, P_{-1,-1}\}$.

\subsubsection{Subleading soft graviton and $\overline{SL(2,\mathbb{C})}$ current algebra} 

The subleading conformal soft graviton with positive helicity is defined as
\begin{equation}
\label{Sp1}
\begin{split}
S_{1}^{+}(z,\bar{z}) = \lim_{\Delta \to 0} \Delta G^{+}_{\Delta}(z,\bar{z}) 
\end{split}
\end{equation}

$S_{1}^{+}(z,\bar{z}) $ can be also be written as \cite{Banerjee:2020zlg}
\begin{equation}
\label{bsl2ccurrents}
\begin{split}
S_{1}^{+}(z,\bar{z})  = - J^{1}(z) + 2 \bar{z} J^{0}(z) - \bar{z}^{2}J^{-1}(z)
\end{split}
\end{equation}

where  $J^{a}(z), a=0,\pm 1$ are currents whose Ward identities are given by the subleading conformal soft graviton theorem. The modes of these currents $J^{a}_{m}$ with $m \in \mathbb{Z}$ satisfy the following current algebra
\begin{equation}
\label{JJcom}
\begin{split}
[J^{a}_{m}, J^{b}_{n}] = (a-b)J^{a+b}_{m+n}
\end{split}
\end{equation}

These modes act on a conformal primary $\mathcal{O}_{h,\bar{h}}(z,\bar{z})$ as 
\begin{equation}
\label{sl2cact}
\begin{split}
[J^{a}_{m}, \mathcal{O}_{h,\bar{h}}(z,\bar{z})] =  z^{m} \bar{z}^{a} \left( \bar{z} \partial_{\bar{z}} + (a+1)\bar{h} \right) \mathcal{O}_{h,\bar{h}}(z,\bar{z})
\end{split}
\end{equation}

where $a=0,\pm 1; m\in \mathbb{Z}$.  From \eqref{sl2cact} we see that the zero modes $J^{a}_{0}$ are the global $\overline{SL(2,\mathbb{C})}$ generators
\begin{equation}
\label{J0mode}
\begin{split}
J^{0}_{0} = \bar{L}_{0}, \quad J^{1}_{0} = \bar{L}_{1}, \quad J^{-1}_{0} = \bar{L}_{-1}
\end{split}
\end{equation}

The $m\ne 0$ modes generate local $\overline{SL(2,\mathbb{C})}$ transformations. We will refer to \eqref{JJcom} as the $\overline{SL(2,\mathbb{C})}$ current algebra. 

\subsubsection{Extended symmetry algebra}

The supetranslation generators $P_{n,0}$ and $P_{n,-1}$ form a closed algebra with the $\overline{SL(2,\mathbb{C})}$ current algebra geneators $J^{a}_{n}$. This is  given by
\begin{equation}
\label{JPcom}
\begin{split}
[J^{a}_{m}, P_{n,0}] = \frac{(a-1)}{2} P_{m+n,a}, \quad [J^{a}_{m}, P_{n,-1}] = \frac{(a+1)}{2} P_{m+n,a-1}
\end{split}
\end{equation}

where $a=0,\pm 1$ and $m, n \in \mathbb{Z}$.  

Now the global symmetries of the S-matrix also include the global $SL(2,\mathbb{C})$ generators $\{L_{0}, L_{\pm 1}\}$\footnote{This is due to Lorentz invariance of the S-matrix. The $4$-$d$ Lorentz group is generated by \\ $\{L_{0},L_{\pm 1}, \bar{L}_{0}, \bar{L}_{\pm 1}\}$.}. Their commutators with $J^{a}_{n}, P_{n,0}$  and  $ P_{n,-1} $ are as follows
\begin{equation}
\label{LJLPcom}
\begin{split}
& [L_{m}, J^{a}_{n}] = - n J^{a}_{m+n},   \quad \hspace{2.2cm} m= 0,\pm 1; n \in \mathbb{Z} \\
& [L_{m}, P_{r,s}] = \left(\frac{m-1}{2}-r\right) P_{m+r, s}, \quad m=0,\pm 1; s=0,-1; r \in \mathbb{Z} 
\end{split}
\end{equation}



\subsection{Null states and differential equations for MHV graviton amplitudes}

In \cite{Banerjee:2020zlg} it was shown that in the MHV sector there exist null state relations among certain linear combinations of descendants created by the action of the infinite-dimensional symmetry algebra generators discussed above. Two of these null states which played a crucial role in that paper are\footnote{In the MHV sector the null state relation \eqref{nst1} also holds for a negative helicity graviton primary. But \eqref{nst2} exists only for positive helicity graviton primaries. }
\begin{align} \label{nst1}
& \Psi_{\D} = \left[J^1_{-1} P_{-1,-1} - (\Delta-1) P_{-2,0}\right]G^{+}_{\D}(z,\bar{z}) = 0 
\end{align}

and
\begin{align} 
\label{nst2}
& \Phi_{\Delta} = \left[L_{-1} P_{-1,-1}+ 2 \hspace{0.04cm} J^{0}_{-1} P_{-1,-1}- (\Delta+1) P_{-2,-1} - \bar{L}_{-1}P_{-2,0} \right] G^{+}_{\Delta}(z,\bar{z}) =0
\end{align}


Equations \eqref{nst1} and \eqref{nst2} hold as operator statements only within MHV amplitudes. Both $ \Psi_{\D} $ and $ \Phi_{\D} $ are primaries under supertranslations, the $\overline{SL(2,\mathbb{C})}$ current algebra and global $SL(2,\mathbb{C})$ and are uniquely determined by these symmetries. The existence of these null states is also a consequence of the consistency of the celestial OPE of graviton primaries in the MHV sector with the leading and subleading soft graviton theorems. 

Now the decoupling of these null states lead to partial differential equations (PDEs) for MHV graviton amplitudes. In particular, for an $n$-point MHV amplitude there are two sets of $(n-2)$ linear first order PDEs corresponding to the $(n-2)$ positive helicity gravitons in the MHV amplitude. We refer the reader to Section $15$ of \cite{Banerjee:2020zlg} for explicit expressions of these differential equations. Using these PDEs, the leading term in the celestial OPE of graviton primaries, extracted from MHV amplitudes, can be completely fixed as shown in \cite{Banerjee:2020zlg}.  

Subleading terms corresponding to descendant contributions in the graviton OPEs in the MHV sector were also analysed in \cite{Banerjee:2020zlg}. In particular such terms involve supertranslation and $\overline{SL(2,\mathbb{C})}$ current algebra descendants. The descendant OPE coefficients can be obtained in terms of the leading OPE coefficient by demanding both sides of the celestial OPE to transform in the same fashion under the local symmetry algebra generated by $\{ J^{a}_{n}, P_{n,0}, P_{n,-1}, L_{0}, L_{\pm 1} \}$. This symmetry based derivation of descendant OPEs was carried out explicitly for the first few subleading orders in the OPE in \cite{Banerjee:2020zlg} and the results were also found to be in perfect agreement with the OPE extracted from explicit expressions of MHV amplitudes. 

The $\overline{SL(2,\mathbb{C})}$ current algebra and supertranslation symmetries together with global $SL(2,\mathbb{C})$  therefore allow us to compute the full OPE in the MHV sector in Einstein gravity. Consequently these symmetries  are sufficient to determine the full tree-level MHV graviton scattering amplitudes.

Before ending this review let us note that so far, the symmetry corresponding to the subsubleading soft graviton theorem has not played any role. In the rest of this paper we will study the implications of the constraints imposed by this symmetry in the MHV sector. 



\section{Subsubleading soft graviton}
\label{subsubsoft}

The positive helicity subsubleading conformally soft \cite{Donnay:2018neh} graviton $S^+_2(z,\bar z)$ is defined as,
\be\label{ssc}
S^+_2(z,\bar z) = \lim_{\D\rightarrow -1} (\D +1) G^+_{\D}(z,\bar z)
\ee

where $G^+_{\D}(z,\bar z)$ is a positive helicity graviton primary with conformal dimension $\D$. Now the soft graviton $S^+_2(z,\bar z)$ can be used as in \cite{Banerjee:2020zlg} to define four currents $\{ S^{i}(z), i =0,1,2,3\}$ 
\be
\label{ss2currs}
S^+_2(z,\bar z) = S^0(z) + \bar z S^1(z) + \bar z^2 S^2(z)+ \bar z^3 S^3(z)
\ee 

The Ward identities satisfied by the currents $S^i(z)$ are given by the subsubleading soft graviton theorem written in Mellin space. The reader is referred to Appendix \ref{comms} for the commutation relations among the modes of the currents in \eqref{ss2currs} and also the commutators with the supertranslation and $\overline{SL(2,\mathbb{C})}$ current algebra generators. 

For our purpose it is more convenient to consider the OPE between the currents $S^i(z)$ and a conformal primary $\phi_{h,\bar h}(0,0)$ and it is given by,  
\be\label{ope}
\begin{split}
 S^+_2(z,\bar z)\phi_{h,\bar h}(0,0) & = \sum_{p=1}^\infty z^{p-1} S^0_{-p}\phi_{h,\bar h}(0,0) \\
& + \bar z \( - \frac{\bar h\(2\bar h -1\)}{z} \epsilon \phi_{h-\frac{1}{2},\bar h-\frac{1}{2}}(0,0) + \sum_{p=1}^\infty z^{p-1} S^1_{-p}\phi_{h,\bar h}(0,0)\) \\
&+ \bar z^2 \( \frac{2\bar h}{z} \epsilon \bar\partial\phi_{h-\frac{1}{2},\bar h -\frac{1}{2}}(0,0) + \sum_{p=1}^\infty z^{p-1} S^2_{-p}\phi_{h,\bar h}(0,0)\) \\ 
&+ \bar z^3 \( - \frac{1}{2} \frac{1}{z} \epsilon \bar\partial^2 \phi_{h-\frac{1}{2},\bar h-\frac{1}{2}}(0,0) + \sum_{p=1}^\infty z^{p-1} S^3_{-p}\phi_{h,\bar h}(0,0)\)
\end{split}
\ee

where $\epsilon=\pm1$ depending on whether the primary $\phi_{h,\bar h}$ is outgoing or incoming. The operators $S^i_{-p}\phi_{h,\bar h}$ for $p\in \mathbb Z^+$ are the descendants of the primary $\phi_{h,\bar h}$ created by the subsubleading currents $S^i(z)$. 

Now the subsubleading soft graviton theorem implies that in the MHV sector there are linear relations between the descendants $S^i_{-p}\phi_{h,\bar h}$ and the descendants of supertranslation and $\overline{SL(2,\mathbb C)}$ current algebra \cite{Banerjee:2020zlg}. In the following section we will derive few such relations starting from the OPE between two positive helicity gravitons in the MHV sector.  

\section{Linear relations involving subsubleading current algebra descendants}
\label{subsubrels1}

In the MHV sector the OPE between two positive helicity gravitons is known \cite{Banerjee:2020zlg}. The leading term \cite{Pate:2019lpp,Banerjee:2020zlg} in the OPE, when both the gravitons are outgoing, is given by
\be
G^+_{\D_1}(z,\bar z) G^+_{\D_2}(0,0) \sim - B(\D_1-1,\D_2-1) \frac{\bar z}{z} G^+_{\D_1 + \D_2}(0,0)
\ee
As it stands the leading term is manifestly consistent with the subsubleading conformal soft theorem. To be more precise, we take the subsubleading conformal soft limit on the graviton $G^+_{\D_1}$ in the manner specified in \eqref{ssc} and the result we get is,
\be
S_2^+(z,\bar z) G^+_{\D_2}(0,0) \sim - \frac{1}{2}(\D_2-2)(\D_2-3) \frac{\bar z}{z} G^+_{ \D_2 -1}(0,0)
\ee  

This is consistent with the OPE \eqref{ope} once we recognize that in this case, $2\bar h = \D_2 -2 $. Now we will apply the same procedure to the $\mathcal O(z^0\bar z^0)$ and the $\mathcal O(z^0\bar z^1)$ terms in the OPE. We focus on these two terms only because the supertranslation and the $\overline{SL(2,\mathbb C)}$ current algebra null states which appear \cite{Banerjee:2020zlg} at this order is enough to determine the MHV amplitudes. 

\subsection{$\mathcal O(z^0\bar z^0)$}

In the MHV sector the $\mathcal O(z^0\bar z^0)$ term in the OPE of two positive helicity outgoing gravitons is given by  \cite{Banerjee:2020zlg},
\be
G^+_{\D_1}(z,\bar z) G^+_{\D_2}(0,0) \sim B(\D_1-1, \D_2 -1) P_{-2,0} G^+_{\D_1 + \D_2 -1}(0,0)
\ee

Now we take the subsubleading conformal soft limit $\D_1\rightarrow -1$. In this limit we get
\be
S^+_2(z,\bar z) G^+_{\D_2}(0,0)  \sim \frac{1}{2} (\D_2 -2)(\D_2 -3) P_{-2,0}G^+_{\D_2 -2}(0,0)
\ee

The $\mathcal O(z^0\bar z^0)$ term on the L.H.S of \eqref{ope} can be written as $S^0_{-1}G^+_{\D_2}(0,0)$. So consistency with the subsubleading conformal soft theorem requires the following relation
\be\label{ssn1}
S^0_{-1} G^+_{\D} =  \frac{1}{2} (\D -2)(\D -3) P_{-2,0}G^+_{\D -2}
\ee
where we have renamed $\D_2 \rightarrow \D$. This is the first nontrivial relation between the subsubleading and the supertranslation descendants of a graviton primary.

\subsection{$\mathcal O(z^0\bar{z}^1)$}
The $\mathcal O(z^0\bar z^1)$ term in the OPE of two gravitons is given by  \cite{Banerjee:2020zlg},
\be
\begin{gathered}
G^+_{\D_1}(z,\bar z) G^+_{\D_2}(0,0) \\
\sim B(\D_1-1,\D_2-1) \bar z \(\frac{2(\D_1-1)}{\D_1 + \D_2 -2} J^0_{-1} P_{-1,-1} - \D_1 P_{-2,-1}\) G^+_{\D_1 +\D_2 -1}(0,0)
\end{gathered} 
\ee

Now taking the subsubleading conformal soft limit and comparing with the OPE \eqref{ope} we get the linear relation
\be\label{ssn2}
 S^{1}_{-1} G^{+}_{\Delta} -   \frac{1}{2} (\Delta-2)(\Delta-3) P_{-2,-1} G^{+}_{\Delta-2} + 2(\Delta-2) J^{0}_{-1} P_{-1,-1} G^{+}_{\Delta-2} =0
\ee

The relations \eqref{ssn1} and \eqref{ssn2} have been derived from the subsubleading soft theorem. We will now derive these relations algebraically by using the symmetry algebra and the supertranslation and $\overline{SL(2,\mathbb C)}$ current algebra null states \cite{Banerjee:2020zlg} which appear in the MHV sector.

\section{Algebraic derivation}
\label{subsubrels2}

In order to derive \eqref{ssn1} we start from the null state relation  \cite{Banerjee:2020zlg}
\be\label{1}
\Psi_{\D} = J^1_{-1} G^+_{\D} - (\D-2) P_{-2,0}G^+_{\D-1} = 0
\ee
For our purpose we choose the current $S^1(z)$ and denote the corresponding charge which generates the global subsubleading transformation as $S^1_0$. We also note the following commutation relations,
\be\label{2}
\[ S^1_0, P_{-2,0}\] = J^1_{-1}, \quad \[ S^1_0, J^1_{-1}\] = 3 S^0_{-1}
\ee
The charge $S^1_0$ acts on a positive helicity graviton primary as,
\be\label{3}
\[S^1_0, G^+_{\D}(0,0)\] = - \frac{1}{2} (\D-2)(\D-3)G^+_{\D-1}(0,0)
\ee

Now we apply $S^1_0$ to the null state $\Psi$ and using \eqref{1}, \eqref{2} and \eqref{3} one can easily show that the following relation is obtained,
\be\label{null2}
S^0_{-1} G^+_{\D} =  \frac{1}{2} (\D -2)(\D -3) P_{-2,0}G^+_{\D -2} 
\ee
So the global subsubleading symmetry together with the null state $\Psi_{\D}$ is sufficient to determine the relation \eqref{ssn1}. 

Let us now consider the second relation \eqref{ssn2}. In order to derive this relation we start with the null state  \cite{Banerjee:2020zlg}
\be\label{20}
\Phi_{\Delta} = \left[L_{-1}P_{-1,-1} + 2 \hspace{0.04cm} J^{0}_{-1}P_{-1,-1} - (\Delta+1) P_{-2,-1} - \bar{L}_{-1}P_{-2,0} \right] G^{+}_{\Delta} =0
\ee
which appears in the MHV sector. As in the previous case we can derive \eqref{ssn2} by applying $S^{1}_{0}$ to the null state \eqref{20}. Using the commutation relations
\begin{equation}
\begin{split}
&  [ S^{1}_{0}, L_{-1}] =  0 , \quad [ S^{1}_{0}, J^{0}_{-1}] = \frac{1}{2} \hspace{0.03cm} S^{1}_{-1} , \quad  [ S^{1}_{0}, P_{-2,-1}] = -2 \hspace{0.03cm}J^{0}_{-1}, \\
&  [ S^{1}_{0}, \bar{L}_{-1}] = -2\hspace{0.03cm} S^{2}_{0}, \quad [ S^{1}_{0}, P_{-2,0}] = J^{1}_{-1}
\end{split}
\end{equation}

and the action of the subsubleading generators $S^{1}_{0}$ and $S^{2}_{0}$ on a graviton primary
\begin{equation}
\begin{split}
& [ S^{1}_{0}, G^{+}_{\Delta}(z,\bar{z}) ] =  - \frac{1}{2} \left[ (\Delta-2)(\Delta-3) + 4 (\Delta-2) \bar{z} \partial_{\bar{z}} + 3 \hspace{0.04cm } \bar{z}^{2}\partial^{2}_{\bar{z}}   \right]G^{+}_{\Delta-1}(z,\bar{z}) \\
& [ S^{2}_{0}, G^{+}_{\Delta}(z,\bar{z}) ] =   \frac{1}{2} \left[ 2 (\Delta-2) + 3 \bar{z} \partial^{2}_{\bar{z}}  \right]G^{+}_{\Delta-1}(z,\bar{z})
\end{split} 
\end{equation}

we obtain the following relation
\begin{equation}
\label{nullstordzbar2}
\begin{split}
\[S^{1}_{0},\Phi_{\Delta}\] = -\frac{1}{2} (\Delta-1)(\Delta-2) \Phi_{\Delta-1} - \bar{L}_{-1}\Psi_{\Delta-1} +\Omega_{\Delta+1}
\end{split}
\end{equation}

where $\Omega_{\Delta}$ is given by
\begin{equation}
\label{subsubnstordzbar1}
\begin{split}
\Omega_{\D} = S^{1}_{-1} G^{+}_{\Delta} -   \frac{1}{2} (\Delta-2)(\Delta-3) P_{-2,-1} G^{+}_{\Delta-2} + 2(\Delta-2) J^{0}_{-1} P_{-1,-1} G^{+}_{\Delta-2} 
\end{split}
\end{equation}
Note that $\Omega_{\D}$ is essentially the L.H.S of \eqref{ssn2}.

Now using the fact that $\Phi_{\Delta-1}$ and $\Psi_{\Delta-1}$ are both null states we arrive at
\begin{equation}
\label{subsubnstordzbar2}
\begin{split}
\[S^{1}_{0},\Phi_{\Delta}\]  =0 \implies \Omega_{\Delta+1} = 0  
\end{split}
\end{equation}

Shifting the conformal dimension $\Delta \rightarrow \Delta-1$ then yields $\Omega_{\D}=0$ which is the relation \eqref{ssn2}.

\section{Invariance of OPE under subsubleading soft symmetry }
\label{subsubope1}

In this section, we consider the action of the global symmetry associated to the subsubleading soft graviton theorem on the celestial OPE between graviton primaries. Under this global symmetry, the leading term in the OPE is invariant as shown in \cite{Pate:2019lpp}, where this symmetry was used to constrain the leading OPE coefficient. So, our main focus here will be on the subleading terms in the OPE. For definiteness we consider here the first subleading correction to the OPE between positive helicity outgoing gravitons in the MHV sector. In section \ref{opeinvcheck} of the Appendix, the action of the global subsubleading symmetry on further subleading orders in the OPE have been analysed.

In the MHV sector, the OPE between positive helicity outgoing gravitons is given by \cite{Banerjee:2020zlg}
\begin{equation}
\label{opemhv}
\begin{split}
 & G^{+}_{\Delta_{1}}(z_{1},\bar{z}_{1}) G^{+}_{\Delta_{2}}(z_{2},\bar{z}_{2})  \\
 & = B(\Delta_{1}-1,\Delta_{2}-1)\bigg[ - \frac{\bar{z}_{12}}{z_{12}} \hspace{0.03cm} P_{-1,-1} +  P_{-2,0} \bigg] G^{+}_{\Delta_{1}+\Delta_{2}-1}(z_{2},\bar{z}_{2}) + \cdots 
\end{split}
\end{equation}

where the dots denote  higher order terms. To simplify the ensuing analysis let us set $z_{2}=0,\bar{z}_{2}=0$. Then the OPE \eqref{opemhv} takes the form
\begin{equation}
\label{opemhv1}
\begin{split}
 & G^{+}_{\Delta_{1}}(z_{1},\bar{z}_{1}) G^{+}_{\Delta_{2}}(0,0)  \sim B(\Delta_{1}-1,\Delta_{2}-1)\bigg[ - \frac{\bar{z}_{1}}{z_{1}} \hspace{0.03cm} P_{-1,-1} +  P_{-2,0} \bigg] G^{+}_{\Delta_{1}+\Delta_{2}-1}(0,0)  
\end{split}
\end{equation} 

Now we want to check if the above OPE is invariant under the global subsubleading soft symmetry generated by $S^{1}_{0}$. For this we take the commutator of both sides of \eqref{opemhv1} with the operator $S^{1}_{0}$ which acts on a graviton primary as
\begin{equation}
\label{S10primdef}
\begin{split}
& [ S^{1}_{0}, G^{+}_{\Delta}(z,\bar{z}) ] =  - \frac{1}{2} \left[ (\Delta-2)(\Delta-3) + 4 (\Delta-2) \bar{z} \partial_{\bar{z}} + 3 \hspace{0.04cm } \bar{z}^{2}\partial^{2}_{\bar{z}}   \right]G^{+}_{\Delta-1}(z,\bar{z})
\end{split}
\end{equation}

Using \eqref{S10primdef}, we obtain the action of $S^{1}_{0}$ on the left hand side of \eqref{opemhv1} to yield
\begin{equation}
\label{S10ord1opelhs}
\begin{split}
& [ S^{1}_{0}, G^{+}_{\Delta_{1}}(z,\bar{z}) G^{+}_{\Delta_{2}}(0,0)] \\
 & =  - \frac{1}{2} \left[ (\Delta_{1}-2)(\Delta_{1}-3) + 4 (\Delta_{1}-2) \bar{z} \partial_{\bar{z}} + 3 \hspace{0.04cm } \bar{z}^{2}\partial^{2}_{\bar{z}}   \right]G^{+}_{\Delta_{1}-1}(z,\bar{z}) G^{+}_{\Delta_{2}}(0,0)  \\
& - \frac{1}{2} (\Delta_{2}-2)(\Delta_{2}-3)  G^{+}_{\Delta_{1}}(z,\bar{z}) G^{+}_{\Delta_{2}-1}(0,0) 
\end{split}
\end{equation}

Again using the OPE in equation \eqref{S10ord1opelhs} it can be easily shown that 
\begin{equation}
\label{S10ord1opelhs1}
\begin{split}
  [ S^{1}_{0}, G^{+}_{\Delta_{1}}(z,\bar{z}) G^{+}_{\Delta_{2}}(0,0) ] & \sim - \frac{1}{2} \hspace{0.03cm} (\Delta_{1}+\Delta_{2}-3) B(\Delta_{1}-1,\Delta_{2}-1) \bigg[  - (\Delta_{1}+\Delta_{2}-2)  \frac{\bar{z}_{1}}{z_{1}} \hspace{0.03cm} P_{-1,-1} \\
  &+ (\Delta_{1}+\Delta_{2}-6) P_{-2,0} \bigg] G^{+}_{\Delta_{1}+\Delta_{2}-2}(0,0) 
\end{split}
\end{equation}

Now let us consider the action of $S^{1}_{0}$ on the right hand side of the OPE \eqref{opemhv1}. To evaluate this we use \eqref{S10primdef}  which implies
\begin{equation}
\label{S10primdef1}
\begin{split}
 &  [ S^{1}_{0}, G^{+}_{\Delta_{1}+\Delta_{2}}(0,0) ] =  - \frac{1}{2} (\Delta_{1}+\Delta_{2}-2)(\Delta_{1}+\Delta_{2}-3) G^{+}_{\Delta_{1}+\Delta_{2}-1}(0,0) 
\end{split}
\end{equation}

and the following commutator from section \ref{Pnmcomms} of the Appendix 
\begin{equation}
\label{s10P20com}
\begin{split}
&   [ S^{1}_{0}, P_{-2,0}]= J^{1}_{-1}
\end{split}
\end{equation}

Then using \eqref{S10primdef1} and \eqref{s10P20com},  the action of $S^{1}_{0}$ on the right hand side of \eqref{opemhv1} gives
\begin{equation}
\label{S10ord1operhs}
\begin{split}
&   B(\Delta_{1}-1,\Delta_{2}-1) \bigg[ \frac{1}{2} (\Delta_{1}+\Delta_{2}-2) (\Delta_{1}+\Delta_{2}-3)  \hspace{0.04cm} \frac{\bar{z}_{1}}{z_{1}} \hspace{0.03cm}  P_{-1,-1}  \\ 
& +   J^{1}_{-1} P_{-1,-1}-  \frac{1}{2}  (\Delta_{1}+\Delta_{2}-3)(\Delta_{1}+\Delta_{2}-4)  P_{-2,0} \bigg]  G^{+}_{\Delta_{1}+\Delta_{2}-2}(0,0) 
\end{split}
\end{equation}

Comparing equations \eqref{S10ord1opelhs1} and \eqref{S10ord1operhs} we see that leading $\mathcal{O}(\frac{\bar{z}_{1}}{z_{1}})$ term matches. But the subleading $\mathcal{O}(z_{1}^{0}\bar{z}^{0}_{1})$ terms are not the same. This apparently implies that both sides of the OPE \eqref{opemhv1} do not transform in the same fashion under the action of $S^{1}_{0}$.  However note that the difference between the $\mathcal{O}(z_{1}^{0}\bar{z}^{0}_{1})$ terms in \eqref{S10ord1opelhs1} and \eqref{S10ord1operhs} is given by

\begin{equation}
\label{S10ord1opediff}
\begin{split}
&  B(\Delta_{1}-1,\Delta_{2}-1) \left[ - J^{1}_{-1}P_{-1,-1}  +  (\Delta_{1}+\Delta_{2}-3)  P_{-2,0} \right] G^{+}_{\Delta_{1}+\Delta_{2}-2}(0,0)\\
& = - B(\Delta_{1}-1,\Delta_{2}-1) \hspace{0.05cm} \Psi_{\Delta_{1}+\Delta_{2}-2}
\end{split}
\end{equation}

where $\Psi_{\Delta_{1}+\Delta_{2}-2}$ is the primary descendant given in \eqref{1} with $\Delta \rightarrow (\Delta_{1}+\Delta_{2}-2)$.  Therefore under the action of $S^{1}_{0}$, both sides of the OPE \eqref{opemhv} transform in the same way modulo a null state. The decoupling of this null state from MHV graviton scattering amplitudes ensures that the OPE \eqref{opemhv}, when inserted within a MHV amplitude, is  invariant under the global subsubleading symmetry generated by $S^{1}_{0}$. In the section \ref{opeinvcheck} of the Appendix, we have shown that the $\mathcal{O}(z_{1})$ and $\mathcal{O}(\bar{z}_{1})$ terms in the OPE between positive helicity gravitons are also invariant under the action of $S^{1}_{0}$ upto null states.

Now let us note that besides $S^{1}_{0}$ we also have the global symmetry generators $S^{0}_{0}, S^{2}_{0}$ and $S^{3}_{0}$.  These generators can be obtained from $S^{1}_{0}$ by taking commutators with the global $\overline{SL(2,\mathbb{C})}$ generators $\bar{L}_{\pm1}$ as follows
\begin{equation}
\label{LbS10comms}
\begin{split}
& [S^{1}_{0}, \bar{L}_{1}] = 3\hspace{0.03cm} S^{0}_{0}, \quad [S^{1}_{0}, \bar{L}_{-1}] = -2 \hspace{0.03cm} S^{2}_{0}, \quad [S^{2}_{0}, \bar{L}_{-1}] = -3\hspace{0.03cm} S^{3}_{0}
\end{split}
\end{equation}

The celestial OPE of conformal primary operators is automatically invariant under the action of $\bar{L}_{\pm 1}$. Thus to check for invariance of the OPE under global subsubleading symmetry it is sufficient to consider $S^{1}_{0}$. Then invariance under the action of the $S^{0}_{0}, S^{2}_{0}$ and $S^{3}_{0}$ will follow from the above commutation relations \eqref{LbS10comms}.

\section{Acknowledgements}
SB would like to thank the organizer and the participants of the Celestial Holography workshop in Princeton Center for Theoretical Studies (PCTS) for excellent talks and discussions on related subjects. The work of SB is partially supported by the Science and Engineering Research Board (SERB) grant MTR/2019/000937 (Soft-Theorems, S-matrix and Flat-Space Holography). SSS would like to thank Sayantani Bhattacharyya for very helpful discussions on various topics in QFT. SSS would also like to thank the string theory group at Institute of Physics, Bhubaneswar, India for giving him the opportunity to do a summer project. The work of SSS is supported by DST-Inspire Fellowship of Government of India. SG would like to thank Yasha Neiman for useful discussions. The work of SG is supported by the Quantum Gravity Unit of the Okinawa Institute of Science and Technology Graduate University (OIST), Japan. 

\section{Appendix}

\appendix




\section{Commutation relations}
\label{comms}

In this section we note various commutation relations involving the subsubleading soft graviton symmetry generators. 

\subsection{Subsubleading currents}

The modes $S^{a}_{n}$, with $a=0,1,2,3$ and $ n\in \mathbb{Z}$, of the subsubleading currents act on a conformal primary operator $\phi_{h,\bar{h}}(z,\bar{z})$ as follows
\begin{align}
    &\left[ S^{0}_{n},\phi_{h,\Bar{h}}(z,\Bar{z}) \right] = \frac{\kappa}{4} \hspace{0.03cm} z^{n}\left[ 2\Bar{h}(2\Bar{h}-1)\Bar{z} + 4\Bar{z}^{2}\Bar{h}\Bar{\partial} + \Bar{z}^{3}\Bar{\partial}^{2} \right] \phi_{h-\frac{1}{2},\Bar{h}-\frac{1}{2}}(z,\Bar{z})\\
    &\left[ S^{1}_{n},\phi_{h,\Bar{h}}(z,\Bar{z}) \right] = -\frac{\kappa}{4} \hspace{0.03cm} z^{n}\left[ 2\Bar{h}(2\Bar{h}-1) + 8\Bar{z}\Bar{h}\Bar{\partial} + 3\Bar{z}^{2}\Bar{\partial}^{2} \right] \phi_{h-\frac{1}{2},\Bar{h}-\frac{1}{2}}(z,\Bar{z})\\
    &\left[ S^{2}_{n},\phi_{h,\Bar{h}}(z,\Bar{z}) \right] = \frac{\kappa}{4} \hspace{0.03cm} z^{n}\left[ 4\Bar{h} \hspace{0.03cm}\Bar{\partial} + 3\Bar{z}\Bar{\partial}^{2} \right] \phi_{h-\frac{1}{2},\Bar{h}-\frac{1}{2}}(z,\Bar{z})\\
    &\left[ S^{3}_{n},\phi_{h,\Bar{h}}(z,\Bar{z}) \right] = -\frac{\kappa}{4} \hspace{0.03cm} z^{n} \hspace{0.03cm} \Bar{\partial}^{2} \phi_{h-\frac{1}{2},\Bar{h}-\frac{1}{2}}(z,\Bar{z})
\end{align}

where $\kappa= \sqrt{32\pi G_{N}}$ and $\bar{\partial}= \frac{\partial}{\partial \bar{z}}$.   

\subsection{Commutators with supertranslation generators}
\label{Pnmcomms}

\begin{align}
   \left[ S^{0}_{m}, P_{n,-1} \right] &= \frac{\kappa}{2} \hspace{0.03cm} J_{n+m+1}^{1},  \hspace{1.5cm} \left[ S^{0}_{m}, P_{n,0} \right] = 0 \\
     \left[ S^{1}_{m}, P_{n,-1} \right] &= -\kappa \hspace{0.03cm} J_{n+m+1}^{0},  \hspace{1.4cm}  \left[ S^{1}_{m}, P_{n,0} \right] = \frac{\kappa}{2} \hspace{0.03cm} J_{n+m+1}^{1} \\
   \left[ S^{2}_{m}, P_{n,-1} \right] &= \frac{\kappa}{2} \hspace{0.03cm} J^{-1}_{n+m+1}, \hspace{1.6cm} \left[ S^{2}_{m}, P_{n,0} \right] = - \kappa \hspace{0.03cm}  J_{n+m+1}^{0}\\
   \left[ S^{3}_{m}, P_{n,-1} \right] &= 0, \hspace{3.0cm} \left[ S^{3}_{m}, P_{n,0} \right] = \frac{\kappa}{2}J^{-1}_{n+m+1}
\end{align}

\subsection{Commutators with $\overline{SL(2,\mathbb{C})}$ current algebra generators}
\label{calgcomms}

\begin{align}
  &  \left[ S_{m}^{0},J^{-1}_{n} \right] = -S^{1}_{m+n}, \quad \left[ S_{m}^{0},J^{0}_{n} \right] = \frac{3}{2} S^{0}_{m+n}, \quad \left[ S_{m}^{0},J^{1}_{n} \right] = 0   \\
  &  \left[ S_{m}^{1},J^{-1}_{n} \right] = -2  S^{2}_{m+n}, \quad  \left[ S_{m}^{1},J^{0}_{n} \right] = \frac{1}{2} S^{1}_{m+n}, \quad \left[ S_{m}^{1},J^{1}_{n} \right] = 3 S^{0}_{m+n} \\
  &  \left[ S_{m}^{2},J^{-1}_{n} \right] = -3S^{3}_{m+n}, \quad \left[ S_{m}^{2},J^{0}_{n} \right] = -\frac{1}{2}S^{2}_{m+n}, \quad \left[ S_{m}^{2},J^{1}_{n} \right] = 2 S^{1}_{m+n} \\
  &  \left[ S_{m}^{3},J^{-1}_{n} \right] = 0, \quad  \left[ S_{m}^{3},J^{0}_{n} \right] = -\frac{3}{2}S^{3}_{m+n}, \quad \left[ S_{m}^{3},J^{1}_{n} \right] =  S^{2}_{m+n}
\end{align}

\subsection{Commutators with global $SL(2,\mathbb{C})$ generators}
\label{Lncomms}

\begin{align}
  &  \left[ L_{m}, S_{n}^{a} \right] = - \frac{(m+1+2n)}{2} \hspace{0.04cm} S^{a}_{m+n}
\end{align}

where $m=0, \pm 1$, $a=0,1,2,3$ and $n \in \mathbb{Z}$. 

\subsection{Commutators among subsubleading symmetry generators}

The subsubleading symmetry generators $S^{0}_{m}, S^{1}_{m}, S^{2}_{m}$ and  $S^{3}_{m}$ do not form a closed algebra. However the following relations holds when the commutators involving these generators act on a conformal primary operator $\phi_{h,\bar{h}}$.  
\begin{align}
    \left[ S^{0}_{m},S^{1}_{n} \right]\phi_{h,\Bar{h}}(0,0)&=-2\kappa \hspace{0.03cm} \Bar{h} \hspace{0.03cm} S^{0}_{m+n}\phi_{h-\frac{1}{2},\Bar{h}-\frac{1}{2}}(0,0)\\
    \left[ S^{0}_{m},S^{2}_{n} \right]\phi_{h,\Bar{h}}(0,0)&=-\kappa\hspace{0.03cm} \Bar{h} \hspace{0.03cm} S^{1}_{m+n}\phi_{h-\frac{1}{2},\Bar{h}-\frac{1}{2}}(0,0)\\
    \left[ S^{0}_{m},S^{3}_{n} \right]\phi_{h,\Bar{h}}(0)&=-\frac{\kappa\Bar{h}}{2} \hspace{0.03cm} S^{2}_{m+n}\phi_{h-\frac{1}{2},\Bar{h}-\frac{1}{2}}(0,0)\\
    \left[ S^{1}_{m},S^{2}_{n} \right]\phi_{h,\Bar{h}}(0,0)&=-\frac{3\kappa\Bar{h}}{2} \hspace{0.03cm} S^{2}_{m+n}\phi_{h-\frac{1}{2},\Bar{h}-\frac{1}{2}}(0,0)\\
       \left[ S^{1}_{m},S^{3}_{n} \right]\phi_{h,\Bar{h}}(0,0)&=-3  \hspace{0.03cm} \kappa \hspace{0.03cm} \Bar{h} \hspace{0.03cm} S^{3}_{m+n}\phi_{h-\frac{1}{2},\Bar{h}-\frac{1}{2}}(0,0) \\
        \left[ S^{2}_{m},S^{3}_{n} \right]\phi_{h,\Bar{h}}(0,0)&=\kappa \hspace{0.03cm} \Bar{\partial} \hspace{0.03cm} S^{3}_{m+n}\phi_{h-\frac{1}{2},\Bar{h}-\frac{1}{2}}(0,0)
\end{align}

\textbf{Note} : In the main part of the paper we have set $\kappa = \sqrt{32\pi G_N}=2$.

\section{Subsubleading null state}
\label{subsubrels3}

In this section we obtain another null state relation involving modes of the subsubleading currents. For this, let us consider the $\mathcal{O}(z)$ terms in the OPE between positive helicity gravitons $G^{+}_{\Delta_{1}}(z,\bar{z})$ and $G^{+}_{\Delta_{2}}(0)$ \footnote{Here we have use the convenient notation $G^+_{\D}(0)\equiv G^+_{\D}(0,0)$.}. In the MHV sector this is given by \cite{Banerjee:2020zlg} 
 \begin{equation}
\label{ordzopemhv}
\begin{split}
& G^{+}_{\Delta_{1}}(z,\bar{z}) G^{+}_{\Delta_{2}}(0)\Big|_{\mathcal{O}(z)}  \\
& =  z \hspace{0.04cm} B(\Delta_{1}-1,\Delta_{2}-1) \left[  \Delta_{1} P_{-3,0} -\frac{(\Delta_{1}-1)}{(\Delta_{1}+\Delta_{2}-2)} \hspace{0.03cm} J^{1}_{-2}P_{-1,-1}  \right]G^{+}_{\Delta_{1}+\Delta_{2}-1}(0) 
\end{split}
\end{equation}

Taking the subsubleading conformal soft limit $\Delta_{1}\rightarrow -1$ in \eqref{ordzopemhv} we get
\begin{equation}
\label{ordzopesqsoft}
\begin{split}
S^{+}_{2}(z,\bar{z}) G^{+}_{\Delta_{2}}(0) & \supset  z \hspace{0.04cm}   \frac{(\Delta_{2}-2)(\Delta_{2}-3)}{2}\left[  -P_{-3,0} +\frac{2}{(\Delta_{2}-3)}J^{1}_{-2}P_{-1,-1}  \right]G^{+}_{\Delta_{2}-2}(0) 
\end{split}
\end{equation}

But according to the subsubleading soft graviton theorem we should get at this order
\begin{equation}
\label{ordzopesqsoft1}
\begin{split}
S^{+}_{2}(z,\bar{z}) G^{+}_{\Delta_{2}}(0)\Big|_{\mathcal{O}(z)} =  S^{0}_{-2}G^{+}_{\Delta_{2}}(0)
\end{split}
\end{equation}

Comparing \eqref{ordzopesqsoft} and \eqref{ordzopesqsoft1} we obtain 
\begin{equation}
\label{ordznst}
\begin{split}
 S^{0}_{-2}G^{+}_{\Delta_{2}}(0) =  \frac{(\Delta_{2}-2)(\Delta_{2}-3)}{2}\left[  -P_{-3,0} +\frac{2}{(\Delta_{2}-3)}J^{1}_{-2}P_{-1,-1}  \right]G^{+}_{\Delta_{2}-2}(0) 
\end{split}
\end{equation}

Equation \eqref{ordznst} is a null state relation. In section \ref{subsubopez} we will apply this relation to check for the invariance of the $\mathcal{O}(z)$ terms in the OPE under the action of global subsubleading symmetry. For this purpose it is convenient to write the above null relation as
\begin{equation}
\label{subsubnstordz}
\begin{split}
\Upsilon_{\Delta} = S^{0}_{-2} G^{+}_{\Delta} +   \frac{1}{2} (\Delta-2)(\Delta-3) P_{-3,0} G^{+}_{\Delta-2} - (\Delta-2) J^{1}_{-2} P_{-1,-1} G^{+}_{\Delta-2} =0
\end{split}
\end{equation}

\subsection{Deriving \eqref{subsubnstordz} from \eqref{nullstordz} using global subsubleading symmetry }

Consider the following null state which was shown to exist in the MHV sector in \cite{Banerjee:2020zlg}
\begin{equation}
\label{nullstordz}
\begin{split}
\Theta_{\Delta} = \left[L_{-1}P_{-2,0} + J^{1}_{-2}P_{-1,-1} - (\Delta+1) P_{-3,0}  \right] G^{+}_{\Delta} = 0
\end{split}
\end{equation}

We will now show that the null state relation \eqref{subsubnstordz} can be derived from the above null state \eqref{nullstordz} using the global subsubleading generator $S^{1}_{0}$. Using the following commutators given in sections \ref{Pnmcomms}, \ref{calgcomms} and \ref{Lncomms} of the Appendix
\begin{equation}
\begin{split}
&  [ S^{1}_{0}, L_{-1}] =  0, \quad  [ S^{1}_{0}, P_{-2,0}] = J^{1}_{-1} , \quad [ S^{1}_{0}, P_{-3,0}] = J^{1}_{-2} , \quad [ S^{1}_{0}, J^{1}_{-2}] = 3\hspace{0.03cm} S^{0}_{-2}
\end{split}
\end{equation}

and the relation 
\begin{equation}
\label{S10primdef2}
\begin{split}
& [ S^{1}_{0}, G^{+}_{\Delta}(0) ] =  - \frac{1}{2}  (\Delta-2)(\Delta-3) G^{+}_{\Delta-1}(0)
\end{split}
\end{equation}

we get
\begin{equation}
\label{nullstordz1}
\begin{split}
[ S^{1}_{0}, \Theta_{\Delta} ]  = L_{-1}\Psi_{\Delta-1} -\frac{1}{2} (\Delta-2)(\Delta-5) \Theta_{\Delta-1}  + \Upsilon_{\Delta+1}
\end{split}
\end{equation}

where $\Psi_{\Delta-1}$ and $\Theta_{\Delta-1}$ are the null states given by \eqref{1} and \eqref{nullstordz} respectively with $\Delta \rightarrow (\Delta-1)$ and   $\Upsilon_{\Delta+1}$ is 
\begin{equation}
\label{subsubnstordz1}
\begin{split}
\Upsilon_{\Delta+1} = S^{0}_{-2} G^{+}_{\Delta+1} +   \frac{1}{2} (\Delta-1)(\Delta-2) P_{-3,0} G^{+}_{\Delta-1} - (\Delta-1) J^{1}_{-2} P_{-1,-1} G^{+}_{\Delta-1} 
\end{split}
\end{equation}

Since $\Psi_{\Delta-1}$ and $\Theta_{\Delta-1}$ are both null states, we see from \eqref{nullstordz1} that 
\begin{equation}
\label{subsubnstordz2}
\begin{split}
[S^{1}_{0}, \Theta_{\Delta}]  =0 \implies \Upsilon_{\Delta+1} = 0  
\end{split}
\end{equation}

Thus $\Upsilon_{\Delta+1}$ should be a null state. Shifting $\Delta \rightarrow (\Delta-1)$ in \eqref{subsubnstordz1} then gives the null state relation \eqref{subsubnstordz}.


\section{Invariance of OPE under subsubleading symmetry: Further checks}
\label{opeinvcheck}

Here we provide further checks of the invariance of the subleading terms in celestial OPE of gravitons under the global subsubleading symmetry. In particular we consider $\mathcal{O}(\bar{z})$ and $\mathcal{O}(z)$ terms in the OPE between positive helicity gravitons in subsections \ref{subsubopezbar} and \ref{subsubopez} respectively. 
 
\subsection{Action of $S^{1}_{0}$ on $\mathcal{O}(\bar{z})$ terms in OPE}
\label{subsubopezbar}

The $\mathcal{O}(\bar{z})$ terms in the OPE of positive helicity outgoing gravitons in the MHV sector is
\begin{equation}
\label{ordzbarope}
\begin{split}
& G^{+}_{\Delta_{1}}(z,\bar{z}) G^{+}_{\Delta_{2}}(0)\Big|_{\mathcal{O}(\bar{z})} \\
& =  \bar{z} \hspace{0.04cm} B(\Delta_{1}-1,\Delta_{2}-1) \left[ \frac{2(\Delta_{1}-1)}{(\Delta_{1}+\Delta_{2}-2)}J^{0}_{-1}P_{-1,-1} - \Delta_{1} P_{-2,-1} \right]G^{+}_{\Delta_{1}+\Delta_{2}-1}(0) 
\end{split}
\end{equation}

Now let us act with $S^{1}_{0}$ on both sides of \eqref{ordzbarope}.  To evaluate this action on the left hand side,  we use equation \eqref{S10ord1opelhs}, substitute the OPE in it and retain only the $\mathcal{O}(\bar{z})$ terms. This yields,
\begin{equation}
\label{S10ordzbaropelhs}
\begin{split}
 & [ S^{1}_{0}, G^{+}_{\Delta_{1}}(z,\bar{z}) G^{+}_{\Delta_{2}}(0)] \Big|_{\mathcal{O}(\bar{z})}\\
 & = -  \frac{1}{2} \hspace{0.04cm} \bar{z} \hspace{0.04cm} B(\Delta_{1}-1,\Delta_{2}-1)  \bigg[  2 ((\Delta_{1}-2)(\Delta_{1}+1)+(\Delta_{1}-1)(\Delta_{2}-3))  J^{0}_{-1}P_{-1,-1}    \\
& -  (\Delta_{1}+\Delta_{2}-3) ((\Delta_{1}-1)(\Delta_{1}+1)+ \Delta_{1}(\Delta_{2}-3)) P_{-2,-1}  \bigg] G^{+}_{\Delta_{1}+\Delta_{2}-2}(0)
\end{split}
\end{equation}

Then applying $S^{1}_{0}$ to the right hand side of \eqref{ordzbarope} and using \eqref{S10primdef1} and the following commutators from sections \ref{Pnmcomms} and \ref{calgcomms} 
\begin{equation}
\begin{split}
&  [ S^{1}_{0}, P_{-2,-1}] = -2 \hspace{0.03cm} J^{0}_{-1}, \quad  [ S^{1}_{0}, J^{0}_{-1}] = \frac{1}{2} \hspace{0.03cm} S^{1}_{-1}
\end{split}
\end{equation}

we obtain
\begin{equation}
\label{S10ordzbaroperhs}
\begin{split}
&    \bar{z} \hspace{0.04cm} B(\Delta_{1}-1,\Delta_{2}-1) \bigg[  (2\Delta_{1} - (\Delta_{1}-1)(\Delta_{1}+\Delta_{2}-3))J^{0}_{-1}P_{-1,-1} \\
& +  \frac{1}{2}  \Delta_{1} (\Delta_{1}+\Delta_{2}-3)(\Delta_{1}+\Delta_{2}-4)  P_{-2,-1}   +  \frac{(\Delta_{1}-1)}{(\Delta_{1}+\Delta_{2}-2)} S^{1}_{-1}P_{-1,-1}\bigg] G^{+}_{\Delta_{1}+\Delta_{2}-2}(0)
\end{split}
\end{equation}

Now the difference between \eqref{S10ordzbaropelhs} and \eqref{S10ordzbaroperhs} is 
\begin{equation}
\label{S10ordzbaropediff}
\begin{split}
& -  \bar{z} \hspace{0.04cm}  \frac{(\Delta_{1}-1)}{(\Delta_{1}+\Delta_{2}-2)} \hspace{0.04cm} B(\Delta_{1}-1,\Delta_{2}-1) \bigg[  2(\Delta_{1}+\Delta_{2}-2) J^{0}_{-1} P_{-1,-1}  \\
& - \frac{1}{2}  (\Delta_{1}+\Delta_{2}-2)  (\Delta_{1}+\Delta_{2}-3)  P_{-2,-1}  + S^{1}_{-1}P_{-1,-1}  \bigg] G^{+}_{\Delta_{1}+\Delta_{2}-2}(0)\\
& =  - \bar{z} \hspace{0.04cm}  \frac{(\Delta_{1}-1)}{(\Delta_{1}+\Delta_{2}-2)} \hspace{0.04cm} B(\Delta_{1}-1,\Delta_{2}-1)  \hspace{0.05cm} \Omega_{\Delta_{1}+\Delta_{2}}
\end{split}
\end{equation}

where $\Omega_{\Delta_{1}+\Delta_{2}}$ is the null state in \eqref{subsubnstordzbar1} with $\Delta \rightarrow (\Delta_{1}+\Delta_{2})$.  Thus, as in section \ref{subsubope1}, we find that under the action of $S^{1}_{0}$, both sides of the OPE \eqref{ordzbarope} transform in the same way upto a null state. 

\subsection{Action of $S^{1}_{0}$ on $\mathcal{O}(z)$ terms in OPE}
\label{subsubopez}

Let us now check the action of $S^{1}_{0}$ on the $\mathcal{O}(z)$ terms in the OPE of positive helicity outgoing gravitons. The OPE at this order in the MHV sector is given by
\begin{equation}
\label{ordzope}
\begin{split}
& G^{+}_{\Delta_{1}}(z,\bar{z}) G^{+}_{\Delta_{2}}(0)\Big|_{\mathcal{O}(z)}  \\
& =  z \hspace{0.04cm} B(\Delta_{1}-1,\Delta_{2}-1) \left[  \Delta_{1} P_{-3,0} -\frac{(\Delta_{1}-1)}{(\Delta_{1}+\Delta_{2}-2)}J^{1}_{-2}P_{-1,-1}  \right]G^{+}_{\Delta_{1}+\Delta_{2}-1}(0) 
\end{split}
\end{equation}

In this case we have 
\begin{equation}
\label{S10ordzopelhs}
\begin{split}
& [ S^{1}_{0}, G^{+}_{\Delta_{1}}(z,\bar{z}) G^{+}_{\Delta_{2}}(0)] \Big|_{\mathcal{O}(z)}\\
& =   -  \frac{1}{2} \hspace{0.04cm} z \hspace{0.04cm}  B(\Delta_{1}-1,\Delta_{2}-1) \hspace{0.04cm} (\Delta_{1}+\Delta_{2}-\Delta_{3}) \bigg[  \left(( \Delta_{1}-3)(\Delta_{1}-1)+ \Delta_{1}(\Delta_{2}-3)\right) P_{-3,0}    \\
& - \frac{(\Delta_{1}-2)(\Delta_{1}-3)+(\Delta_{1}-1)(\Delta_{2}-3)}{(\Delta_{1}+\Delta_{2}-3)} J^{1}_{-2}P_{-1,-1}  \bigg] G^{+}_{\Delta_{1}+\Delta_{2}-2}(0)
\end{split}
\end{equation}

Then using \eqref{S10primdef1} and the following commutators listed in sections  \ref{Pnmcomms} and \ref{calgcomms} 
\begin{equation}
\begin{split}
&  [ S^{1}_{0}, P_{-3,0}] = J^{1}_{-2}, \quad  [ S^{1}_{0}, J^{1}_{-2}] = 3 \hspace{0.03cm} S^{0}_{-2}
\end{split}
\end{equation}

we obtain the action of $S^{1}_{0}$ on the right hand side of \eqref{ordzope} to be
\begin{equation}
\label{S10ordzoperhs}
\begin{split}
&  z \hspace{0.04cm} B(\Delta_{1}-1,\Delta_{2}-1) \bigg[  \left(\Delta_{1} + \frac{1}{2}(\Delta_{1}-1)(\Delta_{1}+\Delta_{2}-3) \right)J^{1}_{-2}P_{-1,-1} \\
& -  \frac{1}{2}  \Delta_{1} (\Delta_{1}+\Delta_{2}-3)(\Delta_{1}+\Delta_{2}-4)  P_{-3,0} \bigg] G^{+}_{\Delta_{1}+\Delta_{2}-2}(0) \\
& -  z \hspace{0.04cm} B(\Delta_{1}-1,\Delta_{2}-1)  \hspace{0.04cm}   \frac{3(\Delta_{1}-1)}{(\Delta_{1}+\Delta_{2}-2)} \hspace{0.04cm}  S^{0}_{-2} G^{+}_{\Delta_{1}+\Delta_{2}}(0)
\end{split}
\end{equation}

The difference between \eqref{S10ordzopelhs} and \eqref{S10ordzoperhs} is then given by
\begin{equation}
\label{S10ordzopediff}
\begin{split}
&  z \hspace{0.04cm} B(\Delta_{1}-1,\Delta_{2}-1)  \hspace{0.04cm}   \frac{3(\Delta_{1}-1)}{(\Delta_{1}+\Delta_{2}-2)}\bigg[  S^{0}_{-2} G^{+}_{\Delta_{1}+\Delta_{2}}(0) \\
&+  \frac{1}{2}(\Delta_{1}+\Delta_{2}-2)(\Delta_{1}+\Delta_{2}-3) P_{-3,0} G^{+}_{\Delta_{1}+\Delta_{2}-2}(0) -  (\Delta_{1}+\Delta_{2}-2) J^{1}_{-2}P_{-1,-1} G^{+}_{\Delta_{1}+\Delta_{2}-2}(0)\bigg]  \\
& = z \hspace{0.04cm} B(\Delta_{1}-1,\Delta_{2}-1)  \hspace{0.04cm}   \frac{3(\Delta_{1}-1)}{(\Delta_{1}+\Delta_{2}-2)} \Upsilon_{\Delta_{1}+\Delta_{2}}
\end{split}
\end{equation}

where $\Upsilon_{\Delta_{1}+\Delta_{2}}$ is the null state given in \eqref{subsubnstordz} with $\Delta \rightarrow (\Delta_{1}+\Delta_{2})$. The above result thus implies that under the action of the global subsubleading symmetry generator $S^{1}_{0}$ the OPE relation \eqref{ordzope} is invariant upto a null state.


\end{document}